\documentclass{article}
\usepackage[latin1]{inputenc}
\usepackage[T1]{fontenc}
\usepackage{amsmath,amsfonts,amssymb}
\usepackage[francais]{babel}
\usepackage{xspace}

\newtheorem{theorem}{Theorem}

\everymath{\displaystyle}
\title{\bf Jordan Normal and Rational Normal Form Algorithms}
\author{ Bernard Parisse, \hfill Morgane Vaughan
\\ Institut Fourier \\ CNRS-UMR 5582 \\ 100 rue des Maths \\
Universit\'e de Grenoble I\\38402 St Martin d'H\`eres C\'edex}
\date{}
\begin{document}
\maketitle

\begin{abstract}
In this paper, we present a determinist Jordan normal form algorithms 
based on the Fadeev formula~:
\[(\lambda \cdot I-A) \cdot B(\lambda)=P(\lambda) \cdot I\] 
where $B(\lambda)$ is 
 $(\lambda \cdot I-A)$'s 
comatrix and $P(\lambda)$ is $A$'s characteristic polynomial.
This rational Jordan normal form algorithm differs from usual
algorithms since it is not based on the Frobenius/Smith normal form but rather
on the idea already remarked in Gantmacher that the non-zero column
vectors of $B(\lambda_0)$ are eigenvectors of $A$ associated to
$\lambda_0$ for any root $\lambda_0$ of the characteristical polynomial. The
complexity of the algorithm is $O(n^4)$ field operations if we know
 the factorization of the characteristic polynomial
(or $O(n^5 \ln(n))$ operations for a matrix
of integers of fixed size). This algorithm has been implemented
using the Maple and Giac/Xcas computer algebra systems.
\end{abstract}

\section{Introduction}

Let's remember that the Jordan normal form of a matrix is:
$$ A=\left(\begin{array}{cccccccc}
 \lambda_1& 0  & 0 & .&.&.&0&0\\
 ? &\lambda_2 &0&.&.&.&0&0 \\
0 & ? & \lambda_3&.&.&.&0&0\\
 0 & 0 & ?&.&.&.&0&0\\
. & . & .&.&.&.&.&.\\
. & . & .&.&.&.&.&.\\
0 & 0 & 0&.&.&?&\lambda_{n-1}&0\\
0 & 0 & 0&.&.&.&?&\lambda_n\\
\end{array}\right)  $$
where there are 1 or 0 instead of the ?.
It corresponds to a full factorization of the characteristical polynomial.
If the field of coefficients is not algebraically closed, this Jordan
form can only be achieved by adding a field extension. 
The Jordan rational normal form is the best diagonal block form that
can be achieved over the field of coefficients, it corresponds to
the factorization of the characteristic polynomial in irreductible
factors without adding any field extension.

In this paper, we first present a complex Jordan normal form
algorithm. This part does not provide an
improvement per se, but it gives, in a simpler case, a taste of
the rational Jordan Normal form algorithm. More precisely we will present
a similar algorithm that provides a rational normal form maximizing 
the number of 0s. This is not a rational Jordan form since 
the non-diagonal block part does not commute with the block-diagonal part, 
but we show that it is fairly easy to convert it to the rational 
Jordan form.

This algorithm is not based on the Frobenius
form (see e.g. Ozello), and assumes that the
characteristic polynomial can be fully factorized 
(see e.g. Fortuna-Gianni for
rational normal forms corresponding to square-free or other 
partial factorization). It might be combined
with rational form algorithm after the Frobenius
step, but it can be used standalone. It has the same complexity as
other deterministic algorithms (e.g. Steel), 
is relatively easy to implement using basic matrix operations,
and could therefore benefit from parallelism (see also Kaltofen et
al. on this topic).

The algorithm of these articles have been implemented in Maple language, 
they work under Maple V.5 or under Xcas 0.5 in Maple compatibility mode.
They are also natively implemented in Giac/Xcas. Please refer to
section \ref{sec:user_guide} to download these 
implementations. 

\section{The complex normal Jordan form}
\subsection{A simplified case}
Let $A$ be a matrix and $B(\lambda)$ be $(\lambda \cdot I-A)$'s comatrix. 
If every eigenvalue is simple, we consider one: $\lambda_0$. 
Then we can write 
\[(\lambda_0 \cdot I-A) \cdot B(\lambda_0)=P(\lambda_0) \cdot I=0\]
The columns of $B(\lambda_0)$ are $A$ eigenvectors for the eigenvalue 
$\lambda_0$. To have a base of $A$'s characteristic space for the 
eigenvalue $\lambda_0$, we just have to calculate the matrix 
$B(\lambda_0)$ (using Hörner's method for example because $B(\lambda)$ 
is a matrices' polynomial) and to reduce the matrix in columns to find 
one that is not null.\\

Our goal is now to find a similar method when we have higher 
eigenvalues multiplicity.

\subsection{Fadeev Algorithm}
First, we need an efficient method to calculate the matrices polynomial 
$B(\lambda)$. 

Fadeev's algorithm makes it possible to calculate both the 
characteristic polynomial ($P(\lambda)=\det(\lambda I-A)$) 
coefficients ($p_i \ (i=0..n)$) and the matrices coefficients 
$B_i \ (i=0\mbox{ . . }n-1)$ of the matrices polynomial giving 
$(\lambda \cdot I-A)$'s comatrix $B(\lambda)$.

\begin{equation} \label{eq:E}
 (\lambda I -A)B(\lambda)=(\lambda I -A) \sum_{k\leq n-1} B_k \lambda^k
= (\sum_{k\leq n} p_k \lambda^k)I =P(\lambda)I
\end{equation}
By identifying the coefficients of $\lambda$'s powers, we find the 
recurrence relations:
\[ B_{n-1}=p_n I=I, \quad B_k-AB_{k+1}=p_{k+1} I \]
But we still miss a relation between $p_k$ and $B_k$, it is given by the~:
\begin{theorem} (Cohen thm )\\
The derivative of the characteristic polynomial $P'(\lambda)$,
equals the  $(\lambda I-A)$ comatrix trace.
\[ \mbox{tr}(B(\lambda))=P'(\lambda) \]
\end{theorem}
The theorem gives $\mbox{tr}(B_k) = (k+1)p_{k+1} $.
If we take the trace in the recurrence relations above, we find:
\[ \mbox{tr}(B_{n-1})=n p_n, \quad (k+1)p_{k+1} -\mbox{tr}(AB_{k+1})
=np_{k+1} \]
Hence if the field of coefficients is of characteristic 0 (or greater
than $n$) we compute $p_{k+1}$ in function of $B_{k+1}$ and then $B_k$~:
\[ p_{k+1}=\frac{\mbox{tr}(AB_{k+1})}{k+1-n}, 
\quad B_k=AB_{k+1}+p_{k+1} I \]

Let's reorder $P$ and $B$'s coefficients~:
\begin{eqnarray*}
P(\lambda) &=& \lambda^n+p_1\lambda^{n-1}+p_2\lambda^{n-2}...+p_n \\
B(\lambda) &=& \lambda^{n-1}I+\lambda^{n-2}B_1+...+B_{n-1}
\end{eqnarray*}
We have proved that~:
\[ \left\{
\begin{array}{ccc}
A_1=A, & p_1=-\mbox{tr}(A), & B_1=A_1+p_1I \\  
A_2=AB_1, & p_2=-\frac{1}{2}\mbox{tr}(A_2), & B_2=A_2+p_2I \\ 
\vdots & \vdots & \vdots \\
A_k=AB_{k-1}, & p_k=-\frac{1}{k}\mbox{tr}(A_k), & B_k=A_k+p_kI
\end{array}
\right.\]

We can now easily program this algorithm to compute the coefficients 
$B_i$ and $p_i$. The number of operations is $O(n^4)$ field operations
using classical matrix multiplication, or better $O(n^{\omega+1})$ using 
Strassen-like matrix multiplication (for large values of $n$).
For matrices with bounded integers coefficients, the complexity would be
$O(n^5 \ln(n))$ or $O(n^{\omega+2}\ln(n))$ 
since the size of the coefficients of $B_k$ is $O(k\ln(k))$.

{\bf Remark}\\
If the field has non-zero characteristic, $P(\lambda)$ should be
computed first, e.g. using Hessenberg reduction (an $O(n^3)$ field
operations), then $B(\lambda)$ can be computed using Horner
division of $P(\lambda)$ by $\lambda I-A$ (an $O(n^4)$ field operation
using standard matrix multiplication).

\subsection{Jordan cycles}
Jordan cycles are cycles of vectors associated to an eigenvalue 
and giving a basis of the characteristic space.
In a cycle associated to $\lambda_0$, giving a vector $v$ of the cycle, 
you can find the next one by multiplying $(A-\lambda_0 \cdot I)$ by $v$ 
and the sum of the sizes of the cycles associated to an eigenvalue is 
its multiplicity.

For example, if $\lambda_0$ has multiplicity 5, with one 
cycle of length 3 and one of length 2, the block associated to 
$\lambda_0$ in the Jordan basis of the matrix will be~: 
\[ \left(\begin{array}{ccccc}
 \lambda_0 & 0 & 0 & 0 & 0 \\
1 & \lambda_0 & 0 & 0 & 0\\
0 & 1 & \lambda_0 & 0 & 0\\
0 & 0 & 0 & \lambda_0 & 0\\
0 & 0 & 0 & 1 & \lambda_0\\
\end{array}\right) \]

We are looking for vectors giving bases of characteristic spaces associated 
to each eigenvalue of $A$, and these vectors must form Jordan cycles.

\subsection{Taylor expansion and the characteristic space.}
Let ($\lambda _i$, $n_i$) be the eigenvalues counted with their
multiplicities. If the field has characteristic 0, we make a Taylor development at the point 
$\lambda _i$ (cf. equation (\ref{eq:E}) p. \pageref{eq:E})~:
\begin{eqnarray*} 
-P(\lambda )I&=&(A-\lambda I)\left(B(\lambda_i )+ B^1(\lambda _i)(\lambda -
\lambda _i)
+ ... +  B^{n-1}(\lambda_i ) (\lambda -\lambda _i)^{n-1} 
\right) \\
&=& -(\lambda -\lambda _i)^{n_i}
\prod _{j\neq i} (\lambda -\lambda _j)^{n_j} I 
\end{eqnarray*}
where $B^k$ is the $k$-th derivative of $B$ divided by $k!$.

If the characteristic of the field of coefficients is not 0, 
the same expansion holds, since the family
$((\lambda -\lambda _i)^k)_k$ is a basis of the vector space of
polynomials of degree less or equal to $n-1$. 
In this case (but also in the former case), 
the value of $B^k$ can be computed using several
Horner division of $B(\lambda)$ by $\lambda-\lambda_0$.

As $A-\lambda I=A-\lambda _i I - (\lambda -\lambda _i)I$, we have for 
the $n_i$ first powers of $\lambda -\lambda _i$:
\begin{eqnarray} \label{eq:rec1}
(A-\lambda _i I) B(\lambda _i)&=&0\\
(A-\lambda _i I) B^1(\lambda _i)&=&B(\lambda_i )\\
& ... & \\
(A-\lambda _i I) B^{n_i-1} (\lambda _i)  &=& 
B^{n_i-2}(\lambda _i)  \\
(A-\lambda _i I) B^{n_i}(\lambda_i) -  B^{n_i-1}(\lambda_i)
&= &-\prod_{j\neq i}(\lambda _i-\lambda _j)^{n_j} I \label{eq:recn}
\end{eqnarray}

\begin{theorem} \label{th:jordan}
The characteristic space associated to $\lambda _i$ is  equal
to the image of $B^{n_i-1}(\lambda _i)$.
\end{theorem}
{\bf Proof~:}\\
We first show that $B^{n_i-1}(\lambda _i)$'s image is included in 
the characteristic space associated to $\lambda_i$ using the fourth 
equation and the ones before.
Let $v$ be a vector, $v\in Im(B^{n_i-1}(\lambda_i))$, 
then $\exists$ $u$ so that  
$v=B^{n_i-1}(\lambda_i) \cdot u$
\begin{eqnarray*}
(A-\lambda_i \cdot I)^{n_i} \cdot v &=& (A-\lambda_i \cdot I)^{n_i-1} \cdot 
B^{n_i-2}(\lambda_i) \cdot u\\
&=&(A-\lambda_i \cdot I)^{n_i-2} \cdot B^{n_i-3}(\lambda_i) 
\cdot u \\
&.&\\
&.&\\
&.&\\
&=&(A-\lambda_i \cdot I) \cdot B(\lambda_i) \cdot u\\
&=&0
\end{eqnarray*}

Now we want to prove that every vector $v$ in the characteristic space 
is also in $B^{n_i-1}(\lambda _i)$'s image. 
We show it by a recurrence on the smallest integer $m$ verifying
$(A-\lambda _i)^{m}v=0$.

For $m=0$,it's obvious because $v=0$.

Let's suppose the case $m$ is true and prove the case $m+1$.
With the equation $(\ref{eq:recn})$, we just have to show that~:
\begin{eqnarray*}
 w &=& (A-\lambda _i) B^{n_i}(\lambda_i) v\\
&=& B^{n_i-1}(\lambda_i) \cdot v-\prod_{j\neq i}(\lambda_i -
\lambda_j)^{n_j} \cdot v
\end{eqnarray*}
is in $B^{n_i-1}(\lambda _i)$'s image, 
because $B^{n_i-1}(\lambda _i) \cdot v$ is 
in $B^{n_i-1}(\lambda_i)$'s image and thus if we prove 
that $w$ is also in, we'll get that 
$\prod_{j\neq i}(\lambda_i -\lambda_j)^{n_j} \cdot v$ is in and $v$ is
in.

As $B^{n_i}(\lambda_i)$ commutes with $A$ (because it's a polynomial in $A$):
\[ (A-\lambda _i)^m w=B^{n_i}(\lambda_i) (A-\lambda _i)^{m+1}v=
0 \]
We can now apply the recurrence hypothesis to $w$. We now know that 
$w \in B^{n_i-1}(\lambda_i)$. And so does $v$.

\subsection{Algorithm}
To find the Jordan cycles, we apply a Gauss reduction on the columns of 
the matrices $B^{(k)}(\lambda _i)$ 
where $k<n_i$. Doing that at the same time for all the matrices allow us 
to keep the relations (\ref{eq:rec1}) to (\ref{eq:recn}) 
between them after reduction.

Let's think of the matrices one under another, columns aligned. We reduce 
the matrix $B(\lambda _i)$ and we {\em rewrite\/} the elementary operations 
on columns done to $B(\lambda _i)$ on all the matrices 
$B^{k}(\lambda _i)$ to keep the relations between them.

Once the matrix $B(\lambda _i)$ is reduced, if we have $k$ columns not null 
then we know that we already have $k$ $n_i$-long Jordan cycles, 
taking the $k$ chains of $n_i$ vectors associated to the considered column. 
(In fact $k$ is $0$ or $1$ at the first step, see the last paragraph in this
section).

If we don't yet have enough vectors to make a base of the characteristic 
space associated to $\lambda_i$, for each chain of columns of the 
$B^{k}(\lambda _i)$ corresponding to a column of $B(\lambda_i)$ 
that isn't null, we shift by one matrix down all the columns. 
This process keeps the relations between the matrices.

 Again, We reduce, collect the $n_i-1$-long Jordan cycles and shift 
the non-null columns. And again as long as we still need vectors to make 
a basis. 

{\em Remark}:\\
If there are still columns that are not null after the reduction 
of $B(\lambda_i)$, there is really only one because one $n_i$-long cycle 
already gives a base of $\lambda_i$'s characteristic space if $\lambda_i$'s 
multiplicity is $n_i$. If there are more than one cycle associated 
to $\lambda_i$, $B(\lambda_i)$ must be null and we can take care 
of $B^{(1)}(\lambda_i)$, etc..

\subsection{Implementation}
We present here the maple langage implementation.

\subsubsection{Useful functions}
Before implementing this Jordan normal form algorithm, we have implemented
the Fadeev algorithm to calculate the $B_i$ and the characteristic 
polynomial's coefficients, then the Hörner algorithm to calculate the 
$B(\lambda_i)$.
\begin{itemize}
\item  \verb|fadeev(A,Bliste,pliste)|
takes a matrix $A$ and put the $B(\lambda)$ and characteristic
polynomial's coefficients,in \verb|Bliste| and in \verb|pliste| respectively,ordered by increasing powers to make it easier to program a 
polynomial derivation. This step requires $O(n^{\omega+1})$ field operations
($\omega=3$ for classical matrix multiplication).
\item \verb|evalpolymat(l,a)| takes a list of 
matrices, considered as a polynomial's coefficients ordered like 
before, and a number\verb|a|, and gives back the matrix 
calculed by the Hörner's method. Each evaluation requires $O(n^3)$ field
operations (expect $n$ evaluations for a generic matrix with complex
coefficients).
\end{itemize}
Then a few utilities~:
\begin{itemize}
\item \verb|derive_listemat(liste)| takes a list as 
in \verb|evalpolymat| and gives back the derivated list.
\item  \verb|construction_colonneB(Bliste,pliste)| 
takes what is calulated by \verb|fadeev| and gives back a list of 
$p$ lists if the characteristic polynomial has $p$ roots. In each list, 
there is first a couple giving an eigenvalue and its multiplicity and then 
the matrix of the $B^{(k)}(\lambda _i)/k!$ for $k$ from 0 to 
($\lambda_i$'s multiplicity)$ - 1$ stuck one under another.
\item \verb|construction(l,n)| makes the matrix of the eigenvectors using 
a list of eigenvalues and associated cycles. $n$ is the size of the matrix 
we are studying. (see the Algorithm part for more details on 
the list used by this function).
\end{itemize}

The previous section showed that the algorithm requires a reduction 
in columns of the matrix. Maple has a function, called \verb|gaussjord|, 
that makes reduction but in rows, not columns, so after constructing the 
column matrix, we will work with its transposed matrix. 
To work with it, we needed a few more functions~:
\begin{itemize}
\item \verb|test_ligne_nulle(B,i)| takes a matrix 
that doen't have to be square, for example a matrix with $n$ rows 
and $m$ columns (we just have to consider matrices 
where $n\leq m$). 
The function returns $1$ if the $n$  first coefficients of the $i$-th 
rows are null, 0 if not.
\item \verb|decalage_ligne(B,i)| takes the partial $i$-th row
(with $n$ coefficients) and shifts it right by $n$.
\item  \verb|coupe_matrice(B)| 
If the matrix has $n$ 
rows and $m$ columns ( $n\leq m$), this function removes 
the first block $n$x$n$.
\end{itemize}

\subsubsection{The Jordan normal form function}
Splitting the work with all the small functions listed before makes the 
\verb|final| program quite simple. There are three embedded loops,
one loops over all eigenvalues, it constructs the list of matrices
$B^k$ associated to the eigenvalue, the second loop is a while loop
that stops when all characteristic vectors for the current eigenvalue
have been found, the third (inner) loop corresponds to a fixed length of the
cycles that we are finding.

The program creates a list of $p$ lists if the matrix has $p$
eigenvalues that are all different, each of these $p$ lists contains an eigenvalue and the list 
of associated Jordan cycles. Then with the function described above: 
\verb|construction(l,n)| the main program returns the matrix of eigenvectors 
and the Jordan normal form of the matrix $A$.

\subsubsection{Tests matrices}
\[ A=\left(\begin{array}{ccc}
 3 & -1 & 1 \\
2 &0 &1 \\
1 & -1 & 2 
\end{array}\right) \]
\[ B=\left(\begin{array}{ccc}
 3 & 2 & -2 \\
-1 &0 &1 \\
1 & 1 & 0 
\end{array}\right) \]
$A$ has two eigenvalues: 2 (multiplcity 2) and 1 (multiplicity 1). 
$B$ has only one eigenvalue: 1 (multiplicity 3). 
Unlike $A$, the second matrix has two cycles associated to only one 
eigenvalue, it revealed an error in a previous version of the program~: 
in the ``while'' loop, the stop test was inefficient because we could
collect linearly dependent vectors (because the Maple function 
``gaussjord'' making the reduction changes the order of the matrix rows). 
Hence the test function looking if the vector (and the corresponding
Jordan cycle we're about to collect) is independent of the vectors 
already collected 
(by making a matrix with all these vectors and searching the rank).

Once the program showed right for these two examples, it was tested 
successfully on Jordan matrices constructed with JordanBlock and 
BlockDiagonal, moved to another basis by a random matrix conjugation.

\subsubsection{Limits of the implementation}
The first version of the program was not really complete because it
worked only with matrices whose characteristic polynomial, 
``factors'' could factor completely
(e.g. integer matrices with rational eigenvalues, but not integer
matrices with algebraic eigenvalues). 
Since ``solve'' also finds algebraic eigenvalues, 
a ``solve''-answer-like to ``factors''-answer-like
converter was added. Hence this Jordan normal form program is
successfull if and only if ``solve'' is able to find the roots of
the characteristic polynomial.

\section{The Jordan rational normal form}
In the previous section, we sometimes had to introduce an algebraic 
extension of the coefficients field (e.g. $\mathbb{Q}$) 
to be able to compute the characteristic 
polynomial's roots, in this section we will find a basis in the
coefficient field where 
the endomorphism matrix has the best almost diagonal block form, the
{\em Jordan rational normal form}. The diagonal blocks will be
companion matrices (corresponding to irreducible factors of the
characteristical polynomial), and the 1 of the complex Jordan normal
form will be replaced by identity block matrices.

We are first going to compute a normal form with as many zeros as 
possible, and from this form, we will compute the 
Jordan rational form.

\subsection{Pseudo rational Jordan form}
\subsubsection{Algorithm}
The method we're going to use is based on an algorithm similar to the
one used before. Let $Q(\lambda)=q_0+ ...+q_d \cdot \lambda^d$ be an irreducible
factor of the characteristic polynomial in the field of coefficients
of multiplicity $q$ and degree $d$ of the characteristic polynomial $P$. 
Note that $q_d=1$ since $Q$ divides the characteristic polynomial $P$,
hence the euclidean division
algorithm of a polynomial by $Q$ does not require any coefficient
division.

The characteristic space corresponding to the roots of $Q$ will be replaced
by a rational characteristic space of dimension $d \cdot q$ 
made of ``rational Jordan cycles''. 
Recall that~:
\[ (\lambda I -A) \cdot  \sum_{k\leq n-1} B_k \lambda^k=P(\lambda)I \]
Since $Q(\lambda) \cdot I -Q(A)$ is divisible by
$\lambda \cdot I-A$, there exists a matrix $M(\lambda)$ such that~:
\begin{equation} \label{eq:ratjordan}
 (Q(\lambda) I -Q(A)) (\sum_{k\leq n-1} B_k \lambda^k)=
Q(\lambda)^q M(\lambda) 
\end{equation}
Now expand $B(\lambda)$ with respect to increasing powers of
$Q(\lambda)$ by euclidean division by $Q$~: 
\[ B(\lambda)=\sum_k C_k(\lambda) Q(\lambda)^k, 
\quad \mbox{deg}(C_k)<q \] 
Replacing in (\ref{eq:ratjordan}) and observing that the matrix coefficients
of order less than $d$ vanish, we get~:
\[ Q(A) \cdot C_0 = 0, \quad C_k = Q(A) \cdot C_{k+1} \]
This is similar to the case where the eigenvalue is rational, we get a
chain of polynomial matrices that are images of the preceding one by $Q(A)$~:
\[ C_{q-1} \rightarrow C_{q-2} ... \rightarrow C_0 \rightarrow 0 \]
We will find the rational Jordan cycles by constructing Jordan cycles
of $Q(A)$. Note that if we find a Jordan cycle of length $k$ for $Q(A)$ we
can construct $d-1$ other Jordan cycles by multiplying the cycle by
$A^i$ for  $i=1..d-1$. 

All these vectors are independent, indeed if
\[
\sum_{i,j} \lambda_{i,j} A^i Q(A)^j v=0 , \quad Q(A)^k v=0,
Q(A)^{k-1}v \neq 0
\]
by multiplying by $Q(A)^{k-1}$ we get~:
\[
(\sum_i \lambda_{i,k-1} A^i) Q(A)^{k-1} v=0
\]
hence $\lambda_{i,k-1}=0$ for all $i$s since $Q(A)^{k-1} v\neq 0$ and
$Q(A)$ is irreducible. Multiplying further by $Q(A)^{k-2}$, ...,
identity, it follows that all $\lambda_{i,j}$ are zero.

Once we have collected these $k d$ vectors, we search for another cycle
in the vectors of the $C_j$ matrices that are linearly independant to
all $A^i Q(A)^{k-1}v$ starting from $C_0$ and increasing $j$. 
If we find a new
end cycle vector $Q(A)^{k'-1}w$ such that $Q(A)^{k'}w=0$ 
and $Q(A)^{k'-1}w$ is independent of the preceding end-cycle vectors,
then we can form $k'd$ vectors $A^i Q(A)^j w$. We will
show that these vectors are independent of the $A^i Q(A)^j v$ since
$Q(\lambda)=q_0+..+q_d \cdot \lambda_d$ is irreducible. Indeed if
we had a relation like
\[
\sum_{i,j} \lambda_{i,j} A^i Q(A)^j v + \mu_{i,j} A^i Q(A)^j w =0, 
\]
If $j>k'$ then $\lambda_{i,j}=0$ by multiplication by $Q(A)^{j}$
for decreasing $j>k'$.
Now we multiply by $Q(A)^{k'-1}$ and we get two polynomials $P$ and
$R$ of degree less than degree($Q$) such that~:
\[  P(A) Q(A)^{k-1}v + R(A) Q(A)^{k'-1} w=0 \]
Since $Q$ is irreducible, it is prime with $R$ if $R\neq 0$. Hence
if $R\neq 0$, by applying Bézout's
theorem, we could invert $R$ modulo $Q$ and express $w$ as a linear 
combination of $A^i Q(A)^{k-1}v$. Therefore $R=0$ and $P=0$ and
$\mu_{i,k'-1}=\lambda_{i,k'-1}=0$.

Let $(v_{k-1}) \rightarrow (v_{k-2})\rightarrow ... \rightarrow (v_0) 
\rightarrow (0)$ be a cycle of $Q(A)$, we have~:
\[ 
(v_{k-1},Av_{k-1},...,A^{d-1}v_{k-1}) \rightarrow ... \rightarrow
(v_0,Av_0,...,A^{d-1}v_0) \rightarrow (0,...,0)\]
where the arrow means ``image by $Q(A)$''.

Let's write the matrix $A$ in the base 
$v_0,Av_0,..,A^{d-1}v_0,..,v_{k-1},..,A^{d-1}v_{k-1}$~:
we find an  ``almost Jordan rational blockl'', its size is $k \cdot d$~: 
\[ 
\left( \begin{array}{cccccccccc}
0 & 0 & ... & -q_0 &             \ & 0 & 0 & ... & 1 & ... \\
1 & 0 & ... & -q_1 &             \ & 0 & 0 & ... & 0 & ...\\
0 & 1 & ... & -q_2 &             \ & 0 & 0 & ... & 0 & ...\\
\vdots & \vdots & ... & \vdots & \ & \vdots & \vdots & ... & \vdots & ...\\
0 & 0 & ... & -q_{d-1} &         \ & 0 & 0 & ... & 0 & ... \\ 
0 & 0 & ... & 0   &              \ & 0 & 0 & ... & -q_{0} & ... \\
0 & 0 & ... & 0   &              \ & 1 & 0 & ... & -q_{1} & ... \\
\vdots & \vdots & ... & \vdots & \ & \vdots & \vdots & ... & \vdots & ...
\end{array}
\right)
\]
Indeed $v_0$ image by $A$ is $A \cdot v_0$ the second vector basis,
etc. to $A^{d-1} \cdot v_0$ whose image by $A$ is~:
\[ A^d \cdot v_0=(Q(A)-q_0-q_1 \cdot A-...-q_{d-1} \cdot A^{d-1})
\cdot v_0 \]
Since $Q(A) \cdot v_0=0$ ($v_0$ ends a Jordan cycle of $Q(A)$), 
we get the first block of the matrix in the new basis.

For the second block, we get the first $d-1$
columns in a similar way. 
For the last one~:
\[ A^d \cdot v_1=(Q(A)-q_0-q_1 \cdot A-...-q_{d-1} \cdot A^{d-1})
\cdot v_1 \]
Since $Q(A) \cdot v_1=v_0$, we get the above matrix part.
By applying the same method to the rest of the cycle we get the matrix.

\subsubsection{Complexity}
Each euclidean division requires $O(n^3 d)$ field operations
($d$ is the degree of the irreducible factor).
There are $q$ euclidean divisions of a polynomial of degree less than
$n$ with $n,n$ matrices coefficients by a polynomial of degree $d$, 
hence computing the $C_j$ requires
$O(n^3 d q)$ operations, adding for all irreducible factors, we get
a complexity of $O(n^4)$ for the division part.

Let $r_1d, ..., r_qd$ be the number of Jordan cycles of $Q(A)$
of length $q$, ..., 1. We have~:
\[ r_1 q+ r_2 (q-1) + ... + r_q=q\]
The first step of the reduction part requires reducing a $n, n q$ matrix of
rank $r_1 d$. Then we will reduce a $r_1 d+n,n(q-1)$
matrix of rank $(r_1+r_2) d$ such that the $r_1d$ first rows
are already reduced and independant (hence $r_2d$ 
new independent rows in the $n$ last rows remain to be extracted), etc., 
then a $(r_1+...+r_i)d+n,n(q-i)$ matrix of rank
$(r_1+...+r_{i+1})d$ with first $(r_1+...+r_i)d$ independent reduced rows
and $r_{i+1}d$ new independent rows in the
$n$ last rows to extract, etc.
We will have to make $ n r_id$ row operations on the i-th matrix.
Hence we will make $O(n r_i d n (q-i))$ operations on
the i-th matrix. Adding all reduction steps, we will make $O(n^2 d q)$ 
field operations for each irreducible factor, hence $O(n^3)$
field operations for all irreducible factors.

The complexity of the whole pseudo-rational form is therefore 
$O(n^4)$ field operations
and is dominated by the $C_i$ computation (since $B$ can be computed
in $O(n^{\omega+1})$ field operations).

\subsubsection{Example}
\[ A=\left(\begin{array}{ccccccccccc}
1 & & -2 & & 4 & & -2 & & 5 & & -4 \\ \\
0 & & 1&  & \frac{5}{2} & & -\frac{7}{2} & & 2 & & -\frac{5}{2} \\ \\
1 & & -\frac{5}{2} & & 2 & & -\frac{1}{2} & & \frac{5}{2} & & -3 \\ \\
0 & & -1 & & \frac{9}{2} & & -\frac{7}{2} & & 3 & & -\frac{7}{2} \\ \\
0 & & 0 & & 2 & & -2 & & 3 & & -1 \\ \\
1 & & -\frac{3}{2} & & -\frac{1}{2} & & 1 & & \frac{3}{2} & & \frac{1}{2}
\end{array}\right) \]
The characteristic polynomial of  $A$ is $(x-2)^2(x^2-2)^2$. 
For $\lambda=2$ there are 2 eigenvectors~:
\[\left(\begin{array}{cc}
 1 & 0 \\
 \\ 0 & 1\\
 \\ -\frac{26}{9} & -\frac{5}{9}\\
 \\  -\frac{25}{9} &  -\frac{1}{9} \\ 
\\  \frac{55}{9} &  \frac{4}{9}\\
 \\ \frac{53}{9} &  -\frac{4}{9}
\end{array} \right) \]
For $x^2-2$ of multiplicity 2, we find a cycle of length $2$ for 
$Q(A)=A^2-2 \cdot I$:
\[ (0,0,0,-1,-1,-1)\rightarrow(1,0,0,-1,-1,-1)\rightarrow(0,0,0,0,0,0) \]
After multiplication by $A$, we get:
\[ ((0,0,0,-1,-1,-1),(1,4,1,4,0,-3))
\rightarrow((1,0,0,-1,-1,-1),(2,4,2,4,0,-2))
\rightarrow 0 \]
The matrix $P$ is therefore~:
\[ P=\left(\begin{array}{ccccccccccc}
1 & & 2 & & 0 & & 1 & &0 & & 1 \\ \\
0 & & 4&  & 0 & & 4 & & 1 & & 0 \\ \\
0 & & 2 & & 0 & & 1 & & -\frac{5}{9} & &  -\frac{26}{9} \\ \\
-1 & & 4 & & -1 & & 4 & &  -\frac{1}{9} & & -\frac{25}{9} \\ \\
-1 & & 0 & & -1 & & 0 & &  \frac{4}{9} & &  \frac{55}{9} \\ \\
-1 & & -2 & & -1 & & -3 & & -\frac{4}{9} & &  \frac{53}{9}
\end{array}\right) \]
And $A$ becomes~:
\[P^{-1}AP=\left(\begin{array}{cccccc}
0&2&0&1&0&0\\
1&0&0&0&0&0\\
0&0&0&2&0&0\\
0&0&1&0&0&0\\
0&0&0&0&2&0\\
0&0&0&0&0&2
\end{array}\right) \]

To obtain the rational normal form, we must replace
the block $\left(\begin{array}{cc}
0&1\\0&0 \end{array}\right)$
by $\left(\begin{array}{cc}0&1\\1&0 \end{array}\right)$.

\subsection{From pseudo-rational to rational Jordan form}
The pseudo rational form has unfortunately not the commutation
property, 
the block diagonal part does not commute with the remainder, hence
we will compute the rational Jordan form from
the pseudo rational form.

We now assume that we are in a basis where the endomorphism is
in pseudo rational form, and we want to compute a new basis so that
the $\left( \begin{array}{ccc} ... & 0 & 1 \\ ... & 0 & 0 
\\ ... \end{array} \right)$ blocks
are replaced by identity matrices.
Let's assume that we have made the first $j$ blocks (each of size $d$) 
indexed from 0 to $j-1$ corresponding to the family of vectors
$(v_{0,0},...,v_{0,d-1},...,v_{j-1,d-1})$. 
We want to find a vector $v_{j,0}$ to begin the next block.
The $v_{j,l}$ will be defined in function of $v_{j,l-1}$
using the relation $Av_{j,l-1}=v_{j,l}+v_{j-1,l-1}$.
Hence $v_{j,0}$ must satisfy~: 
\begin{equation} \label{eq:jordanrat1}
 Av_{j,d-1}=-q_0 v_{j,0}-...-q_{d-1} v_{j,d-1}+v_{j-1,d-1} 
\end{equation}
Applying the previous recurrence relations, we determine
$Q(A)v_{j,0}$ with respect to $v_{j',l}$ (with
$j'<j$, $l<d$). Since $Q(A)$ is a shift of $d$ indices to the left,
we will let $v_{j,0}$ be the shift of $d$ indices of $Q(A)v_{j,0}$ 
to the right (if we stay in the original basis, ``inverting'' $Q(A)$
can be done using the pseudo-rational basis).

More precisely, let's compute $v_{j,l}$ in terms of the $v_{j,0}$
and $v_{j',l'}$ ($j'<j$). We denote the binomial coefficients by
$\left( ^l_m\right)$ (they can be computed efficiently 
using Pascal's triangle rule). A straightforward recurrence gives~:
\begin{equation} \label{eq:jordanrat3}
v_{j,l} = A^l v_{j,0} - \sum_{m=1}^{\mbox{\small inf}(l,j)} 
\left( ^l _m\right) v_{j-m,l-m}
\end{equation}
Replacing in (\ref{eq:jordanrat1}), we get~:
\[ A^d v_{j,0} - \sum_{m=1}^{\mbox{\small inf}(d,j)} 
\left( ^d _m\right)v_{j-m,d-m}
+ \sum_{l=0}^{d-1} 
q_l (A^l v_{j,0} - \sum_{m=1}^{\mbox{\small inf}(l,j)} \left( ^l _m\right) 
v_{j-m,l-m} )=0
\]
eventually~:
\begin{equation} \label{eq:jordanrat}
 Q(A) v_{j,0}= \sum_{l=1}^d 
q_l \sum_{m=1}^{\mbox{\small inf}(l,j)} \left( ^l _m\right) v_{j-m,l-m} 
\end{equation}

{\bf Application to the example~:}\\
We stay in the original basis for the coordinates.
Here $v_{0,0}=(4,24,12,32,8,-4)$ and $v_{0,1}=Av_{j,0}$. A preimage
by $Q(A)$ is given by $w_{1,0}=(0,4,-4,8,4,-4)$ and $w_{1,1}=Aw_{1,0}$.
Applying (\ref{eq:jordanrat}), and $q_1=0$, $q_2=1$
we must satisfy~:
\[ Q(A) v_{1,0} = \sum_{l=1}^2
q_l \sum_{m=1}^{\mbox{\small inf}(l,1)} \left( ^l _m\right) v_{1-m,l-m} 
 =2v_{0,1} \]
hence ~:
\[\begin{array}{ccccc}
 v_{1,0}&=&2A(0,4,-4,8,4,-4)&=&(-8,-32,0,-48,-16,16) \\
 v_{1,1}&=&Av_{1,0}-v_{0,0}&=&(4,40,-4,64,24,-20) 
\end{array}
\]
We have indeed $Av_{1,1}=2v_{1,0}+v_{0,1}$.

\subsection{Maple implementation}
In the first part, we were working with matrices polynomials and 
not polynomial matrices, so the first thing to do was to create a
\verb|traduction| function (which takes the list of the matrices that 
are $B$'s coefficients, $B$ given by Fadeev algorithm) 
to make the euclidean divisions on $B(\lambda)$ coefficient by
coefficient (\verb|nouvelle_ecriture| function, arguments are $B$
the polynomial $Q$ we want to divide by, and $Q$'s multiplicity). 
Then we collect the cycles of $Q(A)$ as in the complex Jordan form
case, by gluing the $C_i$ matrices vertically and transposing the
result for Gauss-Jordan reductions. The main changes are that we
generate cycles of $A$ by multiplication by $I, A, ..., A^{d-1}$
(\verb|fabriq_cycles| function) and we must take care that a new end-cycle
vector must be independent not only of a previous end-cycle vector
$v_i$ but also of its images $\{A \cdot v_i,..., A^{d-1} \cdot v_i\}$.

The structure of the main rational Jordan form function \verb|Jordan2|
is~:
\begin{itemize}
\item A call to \verb|demarrage| that will return a list
of [[irreducible polynomial, multiplicity],[cycles]].
\item For first order irreducible polynomials, the functions of 
the complex normal form are called
\item For each irreducible polynomial, conversion from pseudo-rational
  Jordan form to rational Jordan form
\item a call to \verb|construction_special| to build the passage matrix.
\end{itemize}

\section{``User guide''} \label{sec:user_guide}
The Giac/Xcas free computer algebra system is available at~:\\
\verb|www-fourier.ujf-grenoble.fr/~parisse/giac.html|\\
The functions \verb|jordan| and \verb|rat_jordan| implement the
Jordan normal form and the rational Jordan normal form.

The maple implementation of this algorithm is available at~:\\
\verb|www-fourier.ujf-grenoble.fr/~parisse/jordan.map|\\
Once the Maple session is opened, run the command \verb|read("jordan.map")|.
Then three programs are available:
\begin{itemize}
\item \verb|TER_Jordan| takes a matrix $A$ and returns the matrix of 
eigenvectors and the Jordan normal form of $A$.
\item \verb|final| takes a matrix $A$ and returns the matrix of
eigenvectors and the pseudo-rational form, calculated with a hybrid 
method combining the two programs above.
\item \verb|Jordan2| takes the matrix $A$ and returns the rational form.
\end{itemize}
Note that in the current version, there is a small inconsistency,
since for the rational roots of the characteristical polynomial,
the Jordan 1 are not on the same side of the diagonal than the Jordan
identity blocs for irreducible factors of degree larger than 1.

This Maple implementation can also be run under Xcas, but it is of
course much faster to call the native Xcas functions.

\section{References}
\begin{itemize}
\item H. Cohen, A Course in Computational Algebraic Number Theory, Springer.
\item Elisabetta Fortuna, Patrizia Gianni
Square-free decomposition in finite characteristic: 
an application to Jordon Form computation, ACM SIGSAM Bulletin, 
v. 33 (4), p. 14-32, 1999   
\item F.R. Gantmacher. The theory of matrices. Chelsea Pub. Co., 
New York, 1959.
\item  Mark Giesbrecht, Nearly Optimal Algorithms For Canonical Matrix
Forms, SIAM Journal on Computing, v.24 n.5, p.948-969, Oct. 1995
\item E. Kaltofen, M.S. Krishnamoorthy, and B.D. Saunders. 
Parallel algorithms for matrix normal forms. 
Linear Algebras and its Appl., 136:189-208, 1990.
\item T.M.L. Mulders, A.H.M. Levelt, normform Maple package, 1993
www.maths.warwick.ac.uk/~bjs/normform 
\item P. Ozello. Calcul exact des formes de Jordan et de Frobenius 
d'une matrice. PhD thesis, Univ. Scientifique et Médicale de Grenoble, 
Grenoble, France, 1987.
\item Allan Steel, A new algorithm for the computation of canonical 
forms of matrices over fields, Journal of Symbolic Computation, 
v.24 n.3-4, p.409-432, Sept./Oct. 1997
\end{itemize}

\end{document}